%% file: cfvs_walcom.tex
\documentclass[11pt]{llncs}

\usepackage{mathrsfs}

\usepackage{vmargin}

\setmarginsrb{1.2in}{1.2in}{1.2in}{1.2in}{0cm}{0.0cm}{0cm}{0.7cm}

\include{preamble}

\begin{document}

\title{FPT-Algorithms for Connected Feedback Vertex Set}

\author{Neeldhara Misra\inst{1} \and Geevarghese Philip\inst{1} \and
Venkatesh Raman\inst{1} \and Saket Saurabh\inst{2} \and Somnath
Sikdar\inst{1}}

\institute{The Institute of Mathematical Sciences, India.\\
\email{\{neeldhara\textbar{}gphilip\textbar{}vraman\textbar{}somnath\}@imsc.res.in}\\
\and The University of Bergen, Norway.\\
\email{saket.saurabh@ii.uib.no} }
\maketitle
\begin{abstract}
We study the recently introduced \textsc{Connected Feedback
Vertex Set (CFVS)} problem from the view-point of parameterized algorithms.
CFVS is the connected variant of the classical 
\textsc{Feedback Vertex Set} problem and is defined as follows: 
given a graph~$G=(V,E)$ and an integer~$k$, decide 
whether there exists~$F\subseteq V$, $|F| \leq k$, 
such that~$G[V \setminus F]$ is a forest 
and~$G[F]$ is connected. We show that \textsc{Connected Feedback Vertex
Set} can be solved in time~$O(2^{O(k)}n^{O(1)})$ on general graphs
and in time~$O(2^{O(\sqrt{k}\log k)}n^{O(1)})$ on graphs excluding
a fixed graph~$H$ as a minor. Our result on general undirected graphs
uses as subroutine, a parameterized algorithm for \textsc{Group Steiner
Tree}, a well studied variant of \textsc{Steiner Tree}. We find the
algorithm for \textsc{Group Steiner Tree} of independent interest
and believe that it will be useful for obtaining parameterized algorithms 
for other connectivity problems. 
\end{abstract}

\section{Introduction}

\textsc{Feedback Vertex Set (FVS)} is a classical NP-complete
problem and has been extensively studied in all subfields of algorithms
and complexity. In this problem we are
given an undirected graph~$G=(V,E)$ and a positive integer~$k$ as input,
and the goal is to check whether there exists a subset~$F\subseteq V$
of size at most~$k$ such that~$G[V\setminus F]$
is a forest. This problem originated in combinatorial circuit design
and found its way into diverse applications such as deadlock
prevention in operating systems, constraint 
satisfaction and Bayesian inference in artificial intelligence. 
We refer to the survey by Festa, Pardalos and Resende~\cite{FestaPR1999}
for further details on the algorithmic study of feedback set problems
in a variety of areas like approximation algorithms, linear programming
and polyhedral combinatorics.

In this paper we focus on the recently introduced connected variant of
\textsc{Feedback Vertex Set}, namely, \textsc{Connected Feedback
Vertex Set (CFVS)}. Here, given a graph~$G=(V,E)$ and a positive
integer~$k$, the objective is to check whether there exists a
vertex-subset~$F$ of size at most~$k$ such that~$G[V\setminus F]$
is a forest and~$G[F]$ is connected. Sitters and Grigoriev~\cite{SittersG2009}
recently introduced this problem and obtained a polynomial time approximation
scheme (PTAS) for CFVS on planar graphs. We find it a bit surprising
that the connected version of FVS was not considered in the literature
until now. This is in complete contrast to the fact that the connected
variant of other problems like \textsc{Vertex Cover}---\textsc{Connected
Vertex Cover}~\cite{EscoffierGM07,FujitoD04,GuoNW07,MolleRR08}, 
\textsc{Dominating Set}---\textsc{Connected 
Dominating Set}~\cite{FominGK08,GuhaK98,GuhaK99,LokshtanovMS09} 
are extremely well studied in the literature. In this
paper, we continue the algorithmic study of CFVS 
from the view-point of parameterized algorithms.

Parameterized complexity is a two-dimensional generalization of {}``P{}
vs.\ NP{}'' where, in addition to the overall input size $n$,
one studies how a secondary measurement that captures additional relevant
information affects the computational complexity of the problem in
question. Parameterized decision problems are defined by specifying
the input, the parameter and the question to be answered. The two-dimensional
analogue of the class~P is decidability within a time bound of $f(k)n^{c}$,
where $n$ is the total input size, $k$ is the parameter, $f$ is
some computable function and $c$ is a constant that does not depend
on $k$ or $n$. A parameterized problem that can be decided in such a
time-bound is termed \textit{fixed-parameter tractable} (FPT). For general
background on the theory see the textbooks by Downey and Fellows~\cite{DowneyF99},
Flum and Grohe~\cite{FlumGrohebook} and Niedermeier~\cite{Niedermeierbook06}.

\name{FVS} has been extensively studied in parameterized algorithms.
The earliest known FPT-algorithms for FVS go back to the late 80's
and the early 90's~\cite{Bodlaender1992,DowneyFellows1992} and used the
seminal Graph Minor Theory of Robertson and Seymour (for an overview,
see~\cite{LovaszGMT}). 
Raman et al.~\cite{RamanSS2002} designed an
algorithm with run-time $O(2^{O(k\log\log k)}n^{O(1)})$ which
basically branched on short cycles in a bounded search tree approach. 
Subsequently, several algorithms for FVS with
run-time~$O(2^{O(k)}n^{O(1)})$ were designed using a technique known
as iterative compression. After several rounds 
of improvements, the current best FPT-algorithm for FVS runs
in time $O(5^{k}kn^{2})$~\cite{ChenFominLiuLuVillanger2007}. 

We show that \name{CFVS} can be solved
in time $O(2^{O(k)}n^{O(1)})$ on general graphs and in time 
$O(2^{O(\sqrt{k}\log k)}n^{O(1)})$ on graphs excluding a fixed
graph~$H$ as a minor. Most of the known
FPT-algorithms for connectivity problems enumerate \emph{all} 
minimal solutions and then try to connect each solution 
using an algorithm for the \textsc{Steiner Tree}
problem. This is the case with the existing FPT-algorithms for \textsc{Connected
Vertex Cover}~\cite{GuoNW07,MolleRR08}. The crucial observation which the algorithms
for \textsc{Connected Vertex Cover} rely on is that there are at 
most~$2^{k}$ minimal vertex covers of size at most~$k$. However, this
approach fails for \name{CFVS}\ as the number of minimal
feedback vertex sets of size at most~$k$ is~$\Omega(n^{k})$
(consider a graph that is a collection of~$k$ vertex-disjoint cycles
each of length approximately~$n/k$). To circumvent this problem, we
make use of ``compact representations'' of feedback vertex sets. A
compact representation is simply a collection of families of mutually
disjoint sets, where each family represents a number of different
feedback vertex sets. This notion was defined by Guo et 
al.~\cite{GuoGrammJCSS2006} who showed that the set
of all minimal feedback vertex sets of size at
most~$k$ can be represented by a collection of set-families of size~$O(2^{O(k)})$.

We use compact representations to obtain an
FPT-algorithm for CFVS in Section~\ref{sec:CFVS_general_graphs}. 
But in order to do that we need  an FPT-algorithm for a 
general version of \textsc{Steiner Tree},
namely \textsc{Group Steiner Tree (GST)}, which is defined as follows. 
Given a graph~$G=(V,E)$, subsets~$T_{i} \subseteq V$,
$1 \leq i \leq l$, and an integer~$p$, does there exist 
a subgraph of~$G$ on~$p$ vertices that is a tree~$T$  
and includes at least one  vertex from each~$T_{i}$. Observe that 
when the $T_{i}$'s are each of size one then GST is the
\textsc{Steiner Tree} problem.

We find it mildly surprising that GST can be solved in time $O(2^{l}n^{O(1)})$
using polynomial space just as \textsc{Steiner Tree} because this is in sharp
contrast to their behaviour in terms of approximability. While {\sc
  Steiner Tree} admits a 1.55-factor approximation 
algorithm~\cite{RobinsZelikovsky2005}, GST is as hard as {\sc Set
  Cover}~\cite{GargKR2000} and hence has an~$\Omega(\log n)$ 
lower bound on the approximation factor~\cite{AlonMS2006}. 
Our FPT-algorithm for GST uses a Turing-reduction to a directed 
version of \textsc{Steiner Tree}, called \textsc{Directed Steiner Out-Tree}, which
we first show to be fixed-parameter tractable. 
We note that GST is known to be of interest to
database theorists, and it has has been studied in \cite{DingYWQZL07}, where
an algorithm with runtime  $O(3^l \cdot n + 2^l \cdot (n + m))$ (that uses
exponential space) is discussed.

We also note that CFVS does not admit a polynomial kernel on general
graphs but has a quadratic kernel on the class of graphs that exclude a fixed
graph~$H$ as minor.
Finally, in Section~\ref{sec:CFVS_H_minor} we design a subexponential time algorithm for CFVS on 
graphs excluding some fixed graph~$H$ as a minor using the theory 
of bidimensionality. This algorithm is obtained using
an $O^{*}(w^{O(w)})$-algorithm that computes an optimal connected
feedback vertex set in graph of treewidth at most~$w$.

\section{Preliminaries}
This section contains some basic definitions related to parameterized
complexity and graph theory as well as commentary on the notation used
in this paper. To describe running times of algorithms we sometimes use the~$O^{*}$
notation. Given~$f: \mathbb{N} \rightarrow \mathbb{N}$, we 
define~$O^{*}(f(n))$ to be~$O(f(n) \cdot p(n))$, where~$p(\cdot)$ is
some polynomial function. That is,
the~$O^{*}$ notation suppresses polynomial factors in the running-time
expression.

A parameterized problem $\Pi$ is a subset of $\Gamma^{*}\times
\mathbb{N}$, where~$\Gamma$ is a finite alphabet. An instance of a 
parameterized problem is a tuple~$(x,k)$, where~$k$ is called 
the parameter. A central notion in parameterized complexity is 
{\em fixed-parameter tractability (FPT)} which means, for a given 
instance~$(x,k)$, decidability in time~$f(k)\cdot p(|x|)$, where~$f$ 
is an arbitrary function of~$k$ and~$p$ is a polynomial in the input 
size. The notion of {\em kernelization} is formally defined as follows. 

\begin{definition}{\rm [\bf Kernelization]} 
A kernelization algorithm, or in short, a kernel for a parameterized problem 
$\Pi\subseteq \Gamma^{*}\times \mathbb{N}$ is an algorithm that given $(x,k)\in \Gamma^{*}\times \mathbb{N} $ 
outputs in time polynomial in $|x|+k$ a pair $(x',k')\in \Gamma^{*}\times \mathbb{N}$ such that (a) $(x,k)\in \Pi$ 
if and only if $(x',k')\in \Pi$ and (b) $|x'|,k'\leq g(k)$, where $g$ is some computable function. The function 
$g$ is referred to as the size of the kernel. If~$g(k)=k^{O(1)}$ or $g(k)=O(k)$ 
then we say that $\Pi$ admits a polynomial kernel and linear kernel respectively. 
\end{definition}

We say that a graph~$G$ (undirected or directed) \emph{contains} a graph~$H$
if~$H$ is a subgraph of~$G$.
Given a directed graph (digraph)~$D=(V,A)$, we let~$V(D)$ and~$A(D)$ denote
the vertex and arc set of~$D$, respectively.  A vertex~$u \in V(D)$ 
is an {\em in-neighbor} ({\em out-neighbor}) of~$v \in V(D)$ 
if~$uv\in A$ ($vu\in A$, respectively). The in- and out-neighborhood 
of a vertex~$v$ are denoted by~$N^{-}(v)$ and~$N^{+}(v)$,
respectively. The {\em in-degree}~$d^-(v)$ 
(resp. {\em out-degree}~$d^+(v)$) of a vertex~$v$ is~$|N^{-}(v)|$ 
(resp. $|N^{+}(v)|$). We say that a subdigraph~$T$ of~$D$ with vertex set~$V_T
\subseteq V(D)$ is an {\em out-tree} if~$T$ is an
oriented tree with only one vertex~$r$ of in-degree zero (called
the {\em root}). The vertices of~$T$ of out-degree zero are called
{\em leaves} and every other vertex is called an {\em internal
  vertex}.

\section{Connected Feedback Vertex in General Graphs}\label{sec:CFVS_general_graphs}
In this section we give an FPT-algorithm for CFVS on general graphs. 
Recall the problem definition.
\parprobdefn{An undirected graph~$G=(V,E)$ and an
integer~$k$.}{The integer~$k$.}{Does there exist~$S\subseteq V$
of size at most~$k$ such that~$G\setminus S$ is acyclic and~$G[S]$ is connected?}
We start by describing an FPT-algorithm for the {\sc Group Steiner Tree} problem 
which is crucially used in our algorithm for CFVS.  

\subsection{Group Steiner Tree}
The \name{Group Steiner Tree (GST)} problem is defined as follows: 
\parprobdefn{An undirected graph~$G=(V,E)$; vertex-disjoint 
subsets~$S_{1},\ldots,S_{l}\subseteq V$; and an integer~$p$.}
{The integer~$l$.}{Does~$G$ contain a tree on at most~$p$ vertices
that includes at least one vertex from each~$S_{i}$?}
Our fixed-parameter algorithm for GST first reduces it
to {\sc Directed Steiner Out-Tree} (defined below) which we then show to 
be fixed-parameter tractable. 
\parprobdefn{A directed graph~$D=(V,A)$; a distinguished 
vertex~$r\in V$; a set of terminals~$S\subseteq V$; and an integer~$p$.} 
{The integer~$l=|S|$.}{Does~$D$ contain an out-tree on at most~$p$ vertices
  that is rooted at~$r$ and that contains all the vertices of~$S$?}

\begin{lemma}\label{lem:gst_dst}
The GST problem Turing-reduces to the \name{Directed Steiner Out-Tree} problem.
\end{lemma}
\begin{proof}
Given an instance~($G=(V,E)$, $S_{1}, \ldots, S_{l},p)$ of GST, 
construct an instance of \name{Directed Steiner Out-Tree} as 
follows. Let~$S = \{s_1, s_2, \ldots , s_l\}$ be a set of~$l$ new vertices, that is,
$s_i \notin V$ for~$1 \leq i \leq l$. Let~$V' = V \cup S$ 
and~$A=\{uv,vu : \{u,v\} \in E\} \cup \bigcup_{i=1}^l \{xs_i : x \in
S_i\}$. Finally, let~$D=(V',A)$. It is easy to see that~$G$ contains a
tree on at most~$p$ vertices that includes at least one vertex from 
each~$S_i$ if and only if there exists a vertex~$r \in V'$ and an 
out-tree in~$D$ rooted at~$r$ on at most~$p+l$ vertices containing all vertices of~$S$.
\qed
\end{proof}

We now show that \textsc{Directed Steiner Out-Tree} is fixed-parameter
tractable. The algorithm outlined here is essentially the same as that for the
\textsc{Steiner Tree} problem due to Nederlof~\cite{Nederlof2009}. We
give an outline for the sake of completeness. 
First recall the well-known Inclusion-Exclusion (IE) formula. 
Let~$U$ be a finite universe and~$A_1, \ldots, A_l
\subseteq U$. Then
\begin{equation}\label{eqn:IE}
\left | \bigcap_{i=1}^{l} A_i\right | = |U| + \sum_{\emptyset \neq X
  \subseteq [l]} (-1)^{|X|} \left| \bigcap_{i \in X} \bar{A}_i\right|.
\end{equation} 
Now note that if for all~$X \subseteq [l]$, one can 
evaluate~$|\bigcap_{i \in X} \bar{A}_i|$ in time polynomial in the
input size~$n$, 
then one can evaluate~$|\bigcap_{i=1}^{l} A_i|$ in time~$O(2^l
\cdot n^{O(1)})$ and using space polynomial in~$n$.

Given a directed graph~$D=(V,A)$, define a \emph{branching walk}~$B$
in~$D$ to be a pair~$(T_B=(V_B,E_B),\phi)$, where~$T_B$ is a rooted ordered
out-tree and~$\phi: V_B \rightarrow V$ is a homomorphism from~$T_B$
to~$D$. The length of~$B$, denoted by~$|B|$, is~$|E_B|$. For a node~$s
\in V$, $B$ is from~$s$ if the root of~$T_B$ is mapped to~$s$
by~$\phi$.  We let~$\phi(V_B)$ denote~$\{\phi(u): u \in V_B\}$
and~$\phi(E_B)$ denote~$\{(\phi(u),\phi(b)): (a,b) \in E_B\}$.  
Let~$(D,r,S,p)$ be an instance of the \textsc{Directed Steiner Tree}
problem. As in~\cite{Nederlof2009}, one can show that there exists 
an out-tree~$T=(V',E')$ of~$D$ rooted at~$r$ such that~$S \subseteq
V'$ and~$|V'| \leq p$ if and only if there exists a
branching walk~$B =(T_B = (V_B,E_B),\phi)$ from~$r$ such that~$S \subseteq
\phi(V_B)$ and~$|B| \leq p-1$. 
We now frame the problem as an
IE-formula. Let~$U$ be the set of all branching walks from~$r$ of
length~$p-1$. For each~$v \in S$, let~$A_v$ be the set of all elements
of~$U$ that contain~$v$. Then~$|\bigcap_{v \in S} A_v|$ is the number of
all branching walks that contain all the vertices of~$S$ and this
number is larger than zero if and only if the instance is a
yes-instance.
 
For~$X \subseteq S$ define~$X' = X \cup (V \setminus S)$, and
define~$b_j^{X}(r)$ to be the number of branching walks from~$r$ of
length~$j$ in the graph~$G[X']$. Then~$|\bigcap_{v \in X} \bar{A}_v| =
b_c^{S \setminus X}(r)$. Now~$b_j^{X}(r)$ can be computed in
polynomial time:  
\[
b_j^{X}(r) = \left \{\begin{array}{ll}
1 & \hspace{1mm} \mbox{if~$j = 0$;} \\ 
\displaystyle{\sum_{s \in N^{+}(r) \cap X'}} \hspace{2mm} \displaystyle{\sum_{j_1 + j_2 = j-1}}
b_{j_1}^{X}(s) \cdot b_{j_2}^{X}(r) & \hspace{1mm} \mbox{otherwise.}
\end{array} \right.
\]
The proof of this again follows from~\cite{Nederlof2009}. Now for
each~$X \subseteq \{1, \ldots, l\}$, we can compute the
term~$|\bigcap_{v \in X} \bar{A}_v|$ in polynomial time and hence
by identity~(\ref{eqn:IE}), we can solve the problem in time~$O(2^{l} \cdot
n^{O(1)})$ using polynomial space.
\begin{lemma}
\label{lem:dst_FPT}
 {\sc Directed Steiner Out-Tree}  can be solved in~$O(2^l \cdot
 n^{O(1)})$ time using polynomial space. 
\end{lemma}

Lemmas~\ref{lem:gst_dst} and~\ref{lem:dst_FPT} together
imply: 
\begin{lemma}
\label{lemma:dsot}
The {\sc Group Steiner Tree} problem can be solved in~$O(2^l
\cdot n^{O(1)})$ time using polynomial space. 
\end{lemma}

One can also solve the  {\sc Group Steiner Tree} problem in 
time $O(2^p \cdot n^{O(1)})$ by reducing it to a weighted 
version of the {\sc Steiner Tree} problem as done in~\cite{MR1192785} and 
then using the polynomial-space algorithm of Nederlof for solving 
{\sc Steiner Tree}~\cite{Nederlof2009}. We believe that {\sc Directed Steiner
  Out-Tree} problem will find various other applications similar 
to the {\sc Steiner Tree} problem.

\subsection{An FPT-Algorithm for CFVS}
Our FPT-algorithm for \name{CFVS}
uses as subroutine an algorithm for enumerating an efficient representation
of minimal feedback vertex sets of size at most~$k$ due to Guo et
al.~\cite{GuoGrammJCSS2006}. Strictly speaking, the subroutine actually
enumerates all compact representations of minimal feedback sets. A
\emph{compact representation} for a set of minimal feedback sets of
a graph~$G=(V,E)$ is a set~$\mathcal{C}$ of pairwise disjoint
subsets of~$V$ such that choosing exactly one vertex from every
set in~$\mathcal{C}$ results in a minimal feedback set for~$G$.
Call a compact representation a \emph{$k$-compact representation}
if the number of sets in the representation is at most~$k$. Clearly,
any connected feedback set of size at most~$k$ must necessarily
pick vertices from the sets of \emph{some} $k$-compact representation.
Given a graph~$G=(V,E)$ and a $k$-compact representation~$S_{1},\ldots,S_{r}$,
where~$r\leq k$, the problem of deciding whether there exists a
connected feedback vertex set with at least one vertex from each set~$S_{i}$
reduces to the \name{Group Steiner Tree} problem where the Steiner
groups are the sets of the compact representation.

Our algorithm therefore cycles through all $k$-compact representations
and for each such representation uses the algorithm for \name{Group Steiner Tree}
to check if there is a tree on at most~$k$ vertices that includes one
vertex from each set~$S_i$ of the compact representation.  If it fails
to do this for all $k$-compact representations, it reports that the
given instance is a \textsc{no}-instance. If it succeeds on some compact
representation, it reports the solution. Since one can enumerate all
compact representations in time~$O(c^{k}\cdot m)$~\cite{GuoGrammJCSS2006}
we have: 
\begin{theorem}
Given a graph~$G=(V,E)$ and an integer~$k$, one can decide whether~$G$
has a connected feedback set of size at most~$k$ in time~$O(c^{k}\cdot n^{O(1)})$,
for some constant~$c$. 
\end{theorem}

Although \name{CFVS} is fixed-parameter tractable, 
it is unlikely to admit a polynomial kernel as the following theorem
shows. This is in contrast to {\sc Feedback Vertex Set} which admits 
a quadratic kernel~\cite{Thomasse2009}. 
\begin{theorem}
The \name{CFVS} problem does not admit a polynomial kernel unless
the Polynomial Hierarchy collapses to~$\Sigma_{3}$. \end{theorem}
\begin{proof}
The proof follows from a polynomial-time parameter-preserving reduction
from \name{Connected Vertex Cover}, which does not admit a polynomial
kernel unless the Polynomial Hierarchy collapses to the third level~\cite{DomLS2009}.
This would prove that \name{CFVS} too does not admit a polynomial
kernel~\cite{BodlaenderDowneyFellowsHermelin2008}. Given an instance~$(G=(V,E),k)$
of the \name{Connected Vertex Cover} problem, construct a new graph~$G'$
as follows: $V(G')=V(G)\cup\{x_{uv}\notin V(G):\{u,v\}\in E(G)\}$;
if~$\{u,v\}\in E(G)$ then add the edges~$\{u,v\},\{u,x_{uv}\},\{x_{uv},v\}$
to~$E(G')$. This completes the construction of~$G'$. It is easy
to see that~$G$ has a connected vertex cover of size at most~$k$
if and only if~$G'$ has a connected feedback vertex set of size
at most~$k$. This completes the proof of the theorem. \qed 
\end{proof}

Interestingly, the results from~\cite{FominLST2010} imply that 
CFVS has polynomial kernel on a graph class~$\cal C$ which excludes 
a fixed graph~$H$ as a minor.

We note in passing that the algorithm for enumerating compact representations can be improved
using results from~\cite{DehneFellowsLRS2005}. The authors of~\cite{DehneFellowsLRS2005}
describe a set of reduction rules such that if a \textsc{yes}-instance
of the \name{Forest Bipartition} problem (defined below) is reduced
with respect to this set of rules then the instance has size at 
most~$5k+1$.
\parnamedefn{Forest Bipartition}{An undirected graph~$G=(V,E)$,
possibly with multiple edges and loops and a set~$S\subseteq V$
such that~$|S|=k+1$ and~$G\setminus S$ is acyclic.}{The integer~$k$.}{Does~$G$
have a feedback vertex set of size at most~$k$ contained in~$V\setminus S$?}
Thus in a \textsc{yes}-instance of \textsc{Forest Bipartition} that
is reduced with respect to the rules in~\cite{DehneFellowsLRS2005},
we have~$|V\setminus S| \leq 4k$. Using this bound in the algorithm
described by Guo et al.~\cite{GuoGrammJCSS2006}, one obtains a $O^{*}(c^{k})$-time
algorithm for enumerating compact representations of minimal feedback
vertex sets of size at most~$k$, where~$c = 52$. The constant~$c$ 
in~\cite{GuoGrammJCSS2006} is more than~$160$. 
\begin{theorem}
{\rm \cite{DehneFellowsLRS2005,GuoGrammJCSS2006}} Given a graph~$G=(V,E)$
and an integer~$k$, the compact representations of all minimal feedback
vertex sets of~$G$ of size at most~$k$ can be enumerated in time~$O(52^{k}\cdot|E|)$. 
\end{theorem}

\section{Subexponential Algorithm for CFVS on $H$-Minor-Free Graphs}
\label{sec:CFVS_H_minor}
In the last section, we obtained an~$O^{*}(c^k)$ algorithm for CFVS
on general graphs.  
In this section we show that CFVS on the class of $H$-minor-free
graphs admits a sub-exponential time algorithm  with 
run-time $O(2^{O(\sqrt{k}\log k)}n^{O(1)})$. 
This section is divided into three parts. In the first part we give 
essential definitions from topological graph theory, the 
second part shows that CFVS can be solved in
time~$O(w^{O(w)}n^{O(1)})$ on graphs with treewidth bounded by~$w$.  
In the last part we give the desired algorithm for CFVS on 
$H$-minor-free graphs by bounding  the treewidth of the input graph using the
known ``grid theorems''. 

\subsection{Definitions and Terminology}
Given an edge~$e$ in a graph~$G$, the \emph{contraction}
of~$e$ is the result of identifying its endpoints in~$G$
and then removing all loops and duplicate edges. 
A {\em minor} of a graph~$G$ is a graph~$H$ that can be obtained from
a subgraph of~$G$ by contracting edges.  
A graph class~$\mathcal C$ is {\em  minor-closed} if any minor of any
graph in~$\mathcal C$ is also an element of~$\mathcal C$. A minor-closed
graph class $\mathcal C$ is {\em $H$-minor-free}  or simply 
{\em $H$-free} if $H \notin \mathcal C$.

A {\em tree decomposition} of a graph~$G=(V,E)$ is a 
pair~$(T=(V_T,E_T), \mathcal{ X}=\{X_{t}\}_{t\in V_T})$ where~$T$ is a tree and
the~$X_t$ are subsets of~$V$ such that:
\begin{enumerate}
\item   $\bigcup_{u\in V_T} X_t =V$;  
\item for each edge~$e = \{u,v\} \in E$ there exists~$t\in V_T$ such
that~$u,v \in X_t$; and 
\item for each vertex~$v \in V$, the subgraph~$T[\{t \mid v \in X_{t}\}]$ is connected. 
\end{enumerate}
The {\em width} of a tree decomposition is $\max_{t\in V_T} |X_t|-1$ and the {\em treewidth} of $G=(V,E)$ is the 
minimum width over all tree decompositions of $G$.

A tree decomposition is called a {\em nice tree decomposition} if the following conditions are satisfied: 
\begin{itemize}
\item Every node of the tree $T$ has at most two children; 
\item if a node $t$ has two children $t_1$ and $t_2$, then $X_t = X_{t_1} = X_{t_2}$; and
\item  if a node $t$ has one child $t_1$, then either $|X_t | =
  |X_{t_1} | + 1$ and $X_{t_1} \subset X_{t}$ or 
  $|X_t| = |X_{t_1} | -1$ and $X_t \subset X_{t_1}$. 
\end{itemize} 
It is possible to transform a given tree decomposition into a nice
tree decomposition in time $O(|V|+|E|)$~\cite{Bodlaender96}.

\subsection{Connected FVS and Treewidth}
In this section we show that the \name{Connected Feedback Vertex Set}
problem is FPT with the treewidth of the input graph as the parameter.
Specifically, we show that the following problem is FPT: 
\parprobdefn{An undirected graph~$G=(V,E)$; an integer~$k$;
and a nice tree decomposition of~$G$ of width~$w$.} {The treewidth~$w$ of 
the graph~$G$.}{Does there exist~$S\subseteq V$ such that
$G\setminus S$ is acyclic, $G[S]$ is connected, and $\left|S\right|\le k$?}

We design a dynamic programming algorithm on the nice tree
decomposition with run-time~$O(w^{O(w)} \cdot n^{O(1)})$ for 
this problem. Let~$\left(T=\left(I,F\right),\left\{ X_{i}\vert i\in I\right\} \right)$ 
be a nice tree decomposition of the input graph~$G$ of width~$w$
and rooted at~$r\in I$. We let~$T_{i}$ denote the subtree of~$T$ 
rooted at~$i\in I$, and~$G_{i}=(V_i,E_i)$ denote the subgraph of~$G$
induced on all the vertices of~$G$ in the subtree~$T_{i}$, that is, $G_{i} = 
G [\bigcup_{j\in V\left(T_{i}\right)}X_{j}]$. 

For each node~$i \in I$ we compute a table~$A_{i}$, the rows of which
are 4-tuples~$[S,P,Y,\mbox{\emph{val}}]$. 
Table~$A_{i}$ contains one row for each combination of the first three
components which denote the following:
\begin{itemize}
\item $S$ is a subset of~$X_{i}$.
\item $P$ is a partition of~$S$ into at most~$\left|S\right|$ labelled
pieces.  
\item $Y$ is a partition of~$X_{i}\setminus S$ into at 
most~$\left|X_i\setminus S\right|$ labelled pieces. 
\end{itemize}
We use~$P(v)$ (resp.~$Y(v)$) to denote the piece of the
partition~$P$ (resp.~$Y$) that contains the vertex~$v$. We let~$|P|$
(resp.~$|Y|$) denote the number of pieces in
the partition~$P$ (resp.~$Y$). 
The last component $\mbox{\emph{val}}$, also denoted as $A_{i}\left[S,P,Y\right]$,
is the size of a smallest feedback vertex set~$F_i \subseteq V\left(G_{i}\right)$
of~$G_i$ which satisfies the following properties:  
\begin{itemize}
\item If~$S=\emptyset$, then~$F_i$ is \emph{connected} in~$G_{i}$.
\item If~$S \ne \emptyset$, then
\begin{itemize}
\item $F_i \cap X_{i} = S$.
\item All vertices of~$S$ that are in any one piece of~$P$ are in a single
connected component of~$G_{i}[F_i]$. Moreover~$G_i[F_i]$ has
exactly~$|P|$ connected components.
\item All vertices of~$X_i \setminus S$ that are in the same piece
  of~$Y$ are in a single connected component (a tree) 
of~$G_{i}[V_i\setminus F_i]$. Moreover~$G_{i}[V_i\setminus F_i]$ has
at least~$|Y|$ connected components.  
\end{itemize}
\end{itemize}
If there is no such set $F_i$, then the last component of the row 
is set to~$\infty$. 

We fix an arbitrary ordering of the vertices of~$X_{i}$, and compute
the table~$A_{i}$ for each node~$i\in I$ of the tree decomposition.
Since there are at most~$w+1$ vertices in each bag~$X_{i}$, there
are no more than 
$$\sum_{i=0}^{w+1} {w+1 \choose i} i^{i} \cdot 
\left(w+1-i\right)^{w+1-i}\leq   \left(2w+2\right)^{2w+2}$$
rows in any table $A_{i}$. 
We compute the tables $A_{i}$ starting
from the leaf nodes of the tree decomposition and going up to the
root. 
\begin{description}
\item [Leaf Nodes.] Let $i$ be a leaf node of the tree decomposition.
We compute the table $A_{i}$ as follows. For each triple $\left(S,P,Y\right)$
where~$S$ is a subset of~$X_{i}$, $P$ a partition of~$S$, 
and~$Y$ a partition of~$X_{i}\setminus S$:

\begin{itemize}
\item Set $A_{i}\left[S,P,Y\right]=\infty$ if at least one of the following
holds:

\begin{itemize}
\item $G_{i}\setminus S$ contains a cycle (i.e., $S$ is \emph{not} an
FVS of $G_{i}$).
\item At least one piece of $P$ is \emph{not} connected in
  $G_{i}[S]$ or if~$G_i[S]$ has less than~$|S|$ connected components.
\item At least one piece of $Y$ is \emph{not} connected 
in~$G_{i}[V_i\setminus S]$ or if~$G_i[V_i \setminus S]$ has less than~$|Y|$ connected components.
\end{itemize}
\item In all other cases, set $A_{i}\left[S,P,Y\right]=\left|S\right|$. 
\end{itemize}
It is easy to see that this computation correctly determines the last
component of each row of $A_{i}$ for a leaf node $i$ of the tree
decomposition.

\item [Introduce Nodes.] Let~$i$ be an introduce node and~$j$ its 
unique child. Let~$x \in X_{i} \setminus X_{j}$
be the introduced vertex. For each triple~$(S,P,Y)$, we compute the 
entry~$A_i[S,P,Y]$ as follows.

\medskip

\noindent {Case~1.} $x \in S$. Check whether~$N(x) \cap S
\subseteq P(x)$; if not, set~$A_i[S,P,Y] = \infty$. 

\begin{itemize}
\item Subcase 1. $P(x) = \{x\}$. Set $A_{i}[S,P,Y] = A_{j}[S \setminus \{x\}, P\setminus P(x),Y]+1$. 
\item Subcase 2: $|P(x)| \ge 2$ and~$N(x) \cap P(x) = \emptyset$.
Set~$A_{i}[S,P,Y] = \infty$, as no extension of~$S$ to an fvs for~$G_i$
can make~$P(x)$ connected. 
\item Subcase 3: $|P(x)| \ge 2$ and~$N(x) \cap P(x) \ne \emptyset$.
Let~$\mathcal{A}$ be the set of all rows~$[S',P',Y]$
of the table~$A_{j}$ that satisfy the following conditions:
\begin{itemize}
\item $S' = S \setminus \{x\}$.
\item $P' = (P \setminus P(x)) \cup Q$, where~$Q$ is a partition
of $P(x) \setminus \{x\} $ such that each piece of~$Q$ contains an 
element of~$N(x) \cap P(x)$.
\end{itemize}
Set~$A_{i}[S,P,Y] = \min_{\mathcal A} \{A_j[S',P',Y]\} + 1$.
\end{itemize}

\medskip

\noindent {Case~2.} $x \notin S$. Check whether~$N(x) \cap (X_i
\setminus S) \subseteq Y(x)$; if not, set~$A_i[S,P,Y] = \infty$. 

\begin{itemize}
\item Subcase 1: $Y(x) = \{x\}$. Set~$A_{i}[S,P,Y] = A_{j}[S,P,Y
  \setminus Y(x)]$.
\item Subcase 2: $|Y(x)| \ge 2$ and~$N(x) \cap Y(x) = \emptyset$.
Set~$A_{i}[S,P,Y] = \infty$, as no extension of~$S$ to an fvs~$F_i$
for~$G_i$ can make~$Y(x)$ a connected component in~$G_i[V_i\setminus F_i]$. 
\item Subcase 3: $|Y(x)| \ge 2$ and~$N(x) \cap Y(x) \ne \emptyset$.
Let~$\mathcal{A}$ be the set of all rows $[S,P,Y']$
of the table $A_{j}$ where~$Y' = (Y \setminus Y(x)) \cup Q$, 
and~$Q$ is a partition of~$Y(x) \setminus \{x\}$ such that 
each piece of~$Q$ contains \emph{exactly}
one element of~$N(x) \cap Y(x)$.
Set~$A_{i}[S,P,Y] = \min_{\mathcal A} \{A_j[S,P,Y']\}$.
\end{itemize}

\item [Forget Nodes.] Let~$i$ be a forget node and~$j$ its 
unique child node. Let~$x \in X_{j} \setminus X_{i}$
be the forgotten vertex. For each triple~$(S,P,Y)$ in the table~$A_i$,
let~$\mathcal{A}$ be the set of all rows~$[S',P',Y]$ of the table~$A_{j}$
that satisfy the following conditions:
\begin{itemize}
\item $S' = S \cup \{x\}$, and
\item $P'(x) = P(y) \cup \{x\}$ for some~$y \in S$. 
\end{itemize}
Let~$\mathcal{B}$ be the set of all rows~$[S,P,Y']$ of the table~$A_{j}$
such that~$Y'(x) = Y(z) \cup \{x\}$ for some~$z \in S$. 
Set \[A_{i}[S,P,Y] = \min \left \{\min_{\mathcal A }A_j[S',P',Y],
\min_{\mathcal B} A_j[S,P,Y'] \right\}.\]

\item [Join Nodes.] Let~$i$ be a join node 
and~$j$ and~$l$ its children. 
For each triple~$(S,P,Y)$ we compute~$A_i[S,P,Y]$ as follows.
\begin{itemize}
\item Case~1. $S = \emptyset$. If both~$A_{j}[\emptyset,P,Y]$
and~$A_{l}[\emptyset,P,Y]$ are positive finite, then 
set $A_{i}[\emptyset,P,Y] = \infty$. Otherwise, 
set~$A_{i}[\emptyset,P,Y] = \max \{ A_{j}[\emptyset,P,Y],
A_{l}[\emptyset,P,Y]\}$.
\item Case~2. $S \ne \emptyset$. Let~$\mathcal{A}$ denote the 
set of all pairs of triples~$\langle (S,P_{1},Y_1),
(S,P_{2},Y_2)\rangle$, where~$(S,P_1,Y_1) \in A_j$ and~$(S,P_2,Y_2)
\in A_l$ with the following property: Starting with the 
partitions~$Q_p = P_{1}$ and~$Q_y = Y_1$ and repeatedly applying 
the following set of operations, we reach stable partitions 
that are identical to~$P$ and~$Y$. 
The first operation that we apply is:
\begin{quote}
If there exist vertices~$u,v \in S$ such that they are in different
pieces of~$Q_p$ but are in the same piece of~$P_{2}$, delete~$Q_p(u)$
and~$Q_p(v)$ from~$Q_p$ and add~$Q_p(u) \cup Q_p(v)$.
\end{quote}

To describe the second set of operations, we need some
notation. Let~$Z = X_i \setminus S$ and let the connected components
of~$G_i[Z]$ be~$C_1, \ldots, C_q$. First contract each connected
component~$C_i$ to a vertex~$c_i$, the \emph{representative} of that
component, and let~$\mathcal{C} = \{c_1, \ldots, c_q\}$. Note that for
each~$1 \le i \le q$, the component~$C_i$ is not split across pieces
in either~$Y_1$ or~$Y_2$. Denote by~$Y_1'$ and~$Y_2'$ the partitions
obtained from~$Y_1$ and~$Y_2$, respectively, be replacing each
connected component~$C_i$ by its representative
vertex~$c_i$. Let~$Q_y = Y_1'$. Repeat until no longer possible:
\begin{quote}
If there exist~$c_a,c_b \in \mathcal{C}$ that are in different pieces
of~$Q_y$ but in the same piece of~$Y_2$ then
delete~$Q_y(c_a),Q_y(c_b)$ from~$Q_y$ and add~$Q_y(c_a) \cup Q_y(c_b)$
provided the following condition holds: for all~$c_e \in \mathcal{C}
\setminus \{c_a, c_b\}$ either~$Y_2(c_e) \cap Q_y(c_a) = \emptyset$
or~$Y_2(c_e) \cap Q_y(c_b) = \emptyset$. 
\end{quote}  
If this latter condition does
not hold, move on to the next pair of triples. Finally expand
each~$c_i$ to the connected component it represents.

Set 
\[A_{i}[S,P,Y] = \min_{\mathcal A} \{A_{j}[S,P_{1},Y_{1}] +
A_{l}[S,P_{2},Y_{2}] - |S|\}.\]

The stated conditions ensure that~$u,v \in S$ are in the same piece
of~$P$ if and only if for each~$\langle (S,P_{1},Y_1),
(S,P_{2},Y_2) \rangle \in \mathcal{A}$, they
are in the same piece of~$P_{1}$ or of~$P_{2}$ (or
both). Similarly, the stated conditions ensure that 
merging solutions at join nodes do not create new cycles.  
Given this, it is easy to verify that the above computation correctly determines
$A_{i}\left[S,P,Y\right]$.
\end{itemize}

\item [Root Node.] We compute the size of a smallest CFVS of~$G$ 
from the table~$A_{r}$ for the root node~$r$ as follows. 
Find the minimum of~$A_{r}[S,P,Y]$
over all triples~$(S,P,Y)$, where~$S \subseteq X_r$, 
$P$ a partition of~$S$ such that~$P$ consists of
a single (possibly empty) piece and~$Y$ is a partition 
of~$X_r\setminus S$. This minimum is the size of a smallest
CFVS of~$G$.
\end{description}

This concludes the description of the dynamic programming algorithm 
for CFVS when the treewidth of the input graph is 
bounded by $w$. From the above description and the size of tables 
being bounded by~$(2w+2)^{2w+2}$, we obtain the following result. 
\begin{lemma}
\label{lemma:treewidth}
Given a graph~$G=(V,E)$, a tree-decomposition of~$G$ of width~$w$, one
can compute the size of an optimum connected feedback vertex set of~$G$ (if it exists)
in time $O((2w+2)^{2w+2} \cdot n^{O(1)})$. 
\end{lemma}

\subsection{FPT Algorithms for $H$-Minor Free Graphs}
We first bound the treewidth of the yes instance of input graphs by $O(\sqrt{k})$. 
\begin{lemma}
If~$(G,k)$ is a yes-instance of CFVS where~$G$ excludes a fixed graph~$H$ as a 
minor, then~$\tw(G) \leq  c_H \sqrt{k}$, where~$c_H$ is a constant
that depends only on the graph~$H$.  
\end{lemma}
\begin{proof}
By~\cite{DemaineH08gra}, for any fixed graph~$H$, every
$H$-minor-free graph~$G$ that does not contain a 
$(w \times w)$-grid as a minor has treewidth at most~$c_H'w$,
where~$c_H'$ is a constant that depends only on the graph~$H$. Clearly
a $(w \times w)$-grid has a feedback vertex set of size at least~$c_1 w^2$,
where~$c_1$ is a constant independent of~$w$. Therefore if~$G$ has a
connected feedback vertex set of size at most~$k$, it cannot have a
$(w \times w)$-grid minor, where~$w > \sqrt{k/c_1}$. Therefore~$\tw(G)
\leq c_H' w \leq c_H' \cdot (\sqrt{k/c_1} +1)  \leq c_H \sqrt{k}$, where~$c_H
= (c_H' + 1)/\sqrt{c_1}$.  
\qed
\end{proof}

\begin{theorem}
\label{thm:subexp}
CFVS can be solved in time~$O(2^{O(\sqrt{k}\log k)}+n^{O(1)})$ on $H$-minor-free graphs.
\end{theorem}
\begin{proof}
Given an instance~$(G,k)$ of CFVS, we first find a tree-decomposition of~$G$ 
using the polynomial-time constant-factor approximation algorithm of 
Demaine et al.~\cite{DemaineHK05}. If $\tw(G) > c_H \sqrt{k}$, then
the given instance is a no-instance; else, 
use Lemma~\ref{lemma:treewidth} to find an optimal CFVS
for~$G$. All this can be done in~$O(2^{O(\sqrt{k} \log k)} \cdot
n^{O(1)})$. To obtain the claimed run-time bound we first apply 
the results from~\cite{FominLST2010} and 
obtain an~$O(k^2)$ kernel for the problem in 
polynomial time and then apply the algorithm described. 
\qed
\end{proof}

\section{Conclusion}
We conclude with some open problems. The obvious question is to obtain
an $O^{*}(c^k)$ algorithm for CFVS in general graphs with a smaller
value of~$c$. Also the approximability of CFVS in general graphs 
is unknown. Is there a constant-factor
approximation algorithm for CFVS? If not, what is the limit of
approximation? Is there an~$O^{*}(c^w)$ algorithm for CFVS, for a
constant~$c$, for graphs
of treewidth at most~$w$? Note that this question is open in the context of
finding a feedback vertex set in graphs of bounded treewidth. 

\bibliographystyle{abbrv}
\bibliography{cfvs}

\end{document}

%% file: preamble.tex
\usepackage{pstricks}
\usepackage{theorem}
\usepackage{amsfonts}
\usepackage{amssymb}
\usepackage{bbm}
\usepackage{dsfont}
\usepackage{euscript}
\usepackage{mathrsfs}
\usepackage{graphics} 
\usepackage{mathrsfs}

\theorembodyfont{\rm}

\newcommand{\name}[1]{\textsc{#1}}

\newcommand{\tw}{{\bf tw}}
\newcommand{\parprobdefn}[3]{
\begin{tabbing}
\emph{Input:} \hspace{1.2cm} \= \parbox[t]{12cm}{#1} \\[1mm]
\emph{Parameter:}            \> \parbox[t]{12cm}{#2} \\
\emph{Question:}             \> \parbox[t]{12cm}{#3} \\
\end{tabbing}
}

\newcommand{\parnamedefn}[4]{
\begin{tabbing}
\name{#1} \\
\emph{Input:} \hspace{1.2cm} \= \parbox[t]{12cm}{#2} \\
\emph{Parameter:}            \> \parbox[t]{12cm}{#3} \\
\emph{Question:}             \> \parbox[t]{12cm}{#4} \\
\end{tabbing}
}



%% file: cfvs_walcom.bbl
\begin{thebibliography}{10}

\bibitem{AlonMS2006}
N.~Alon, D.~Moshkovitz, and S.~Safra.
\newblock Algorithmic construction of sets for {$k$-restrictions}.
\newblock {\em ACM Transactions on Algorithms}, 2(2):153--177, 2006.

\bibitem{Bodlaender1992}
H.~L. Bodlaender.
\newblock On disjoint cycles.
\newblock In G.~Schmidt and R.~Berghammer, editors, {\em Proceedings on
  Graph--Theoretic Concepts in Computer Science ({WG} '91)}, volume 570 of {\em
  LNCS}, pages 230--238. Springer, 1992.

\bibitem{Bodlaender96}
H.~L. Bodlaender.
\newblock A linear-time algorithm for finding tree-decompositions of small
  treewidth.
\newblock {\em SIAM Journal on Computing}, 25(6):1305--1317, 1996.

\bibitem{BodlaenderDowneyFellowsHermelin2008}
H.~L. Bodlaender, R.~G. Downey, M.~R. Fellows, and D.~Hermelin.
\newblock On problems without polynomial kernels (extended abstract).
\newblock In {\em Proceedings of 35th International Colloquium of Automata,
  Languages and Programming {(ICALP 2008)}}, LNCS, pages 563--574. Springer,
  2008.

\bibitem{ChenFominLiuLuVillanger2007}
J.~Chen, F.~V. Fomin, Y.~Liu, S.~Lu, and Y.~Villanger.
\newblock Improved algorithms for the feedback vertex set problems.
\newblock In F.~K. H.~A. Dehne, J.-R. Sack, and N.~Zeh, editors, {\em
  Algorithms and Data Structures, 10th International Workshop, {WADS} 2007,
  Halifax, Canada, August 15-17, 2007, Proceedings}, volume 4619 of {\em
  Lecture Notes in Computer Science}, pages 422--433. Springer, 2007.

\bibitem{DehneFellowsLRS2005}
F.~Dehne, M.~Fellows, M.~A. Langston, F.~Rosamond, and K.~Stevens.
\newblock {An $O(2^{O(k)}n^3)$ FPT-Algorithm for the Undirected Feedback Vertex
  Set problem}.
\newblock In {\em Proceedings of the 11th International Computing and
  Combinatorics Conference (COCOON 2005)}, volume 3595 of {\em LNCS}. Springer,
  2005.

\bibitem{DemaineH08gra}
E.~D. Demaine and M.~Hajiaghayi.
\newblock Linearity of grid minors in treewidth with applications through
  bidimensionality.
\newblock {\em Combinatorica}, 28(1):19--36, 2008.

\bibitem{DemaineHK05}
E.~D. Demaine, M.~Hajiaghayi, and K.~ichi Kawarabayashi.
\newblock Algorithmic graph minor theory: Decomposition, approximation, and
  coloring.
\newblock In {\em Proceedings of the 46th Annual IEEE Symposium on Foundations
  of Computer Science (FOCS 2005)}, pages 637--646. IEEE Computer Society,
  2005.

\bibitem{DingYWQZL07}
B.~Ding, J.~X. Yu, S.~Wang, L.~Qin, X.~Zhang, and X.~Lin.
\newblock Finding top-k min-cost connected trees in databases.
\newblock In {\em ICDE}, pages 836--845. IEEE, 2007.

\bibitem{DomLS2009}
M.~Dom, D.~Lokshtanov, and S.~Saurabh.
\newblock Incompressibility through {C}olors and {ID}s.
\newblock In {\em Proceedings of 36th International Colloquium of Automata,
  Languages and Programming {(ICALP 2009)}}, volume 5555 of {\em LNCS}, pages
  378--389. Springer, 2009.

\bibitem{DowneyFellows1992}
R.~G. Downey and M.~R. Fellows.
\newblock Fixed parameter tractability and completeness.
\newblock In {\em Complexity Theory: Current Research}, pages 191--225.
  Cambridge University Press, 1992.

\bibitem{DowneyF99}
R.~G. Downey and M.~R. Fellows.
\newblock {\em Parameterized Complexity}.
\newblock Springer-Verlag, New York, 1999.

\bibitem{EscoffierGM07}
B.~Escoffier, L.~Gourv{\`e}s, and J.~Monnot.
\newblock Complexity and approximation results for the connected vertex cover
  problem.
\newblock In {\em WG}, volume 4769 of {\em Lecture Notes in Computer Science},
  pages 202--213, 2007.

\bibitem{FestaPR1999}
P.~Festa, P.~M. Pardalos, and M.~G. Resende.
\newblock Feedback set problems.
\newblock In {\em Handbook of Combinatorial Optimization}, pages 209--258.
  Kluwer Academic Publishers, 1999.

\bibitem{FlumGrohebook}
J.~Flum and M.~Grohe.
\newblock {\em Parameterized Complexity Theory}.
\newblock Texts in Theoretical Computer Science. An EATCS Series.
  Springer-Verlag, Berlin, 2006.

\bibitem{FominGK08}
F.~V. Fomin, F.~Grandoni, and D.~Kratsch.
\newblock Solving connected dominating set faster than $2^{n}$.
\newblock {\em Algorithmica}, 52(2):153--166, 2008.

\bibitem{FominLST2010}
F.~V. Fomin, D.~Lokshtanov, S.~Saurabh, and D.~M. Thilikos.
\newblock Bidimensionality and kernels.
\newblock In {\em Proceedings of the 20th Annual ACM-SIAM Symposium on Discrete
  Algorithms (SODA 2010)}, 2010.
\newblock To appear.

\bibitem{FujitoD04}
T.~Fujito and T.~Doi.
\newblock A 2-approximation {NC}-algorithm for connected vertex cover and tree
  cover.
\newblock {\em Information Processing Letters}, 90(2):59--63, 2004.

\bibitem{GargKR2000}
N.~Garg, G.~Konjevod, and R.~Ravi.
\newblock A polylogarithmic approximation algorithm for the group steiner tree
  problem.
\newblock In {\em Journal of Algorithms}, pages 253--259, 2000.

\bibitem{GuhaK98}
S.~Guha and S.~Khuller.
\newblock Approximation algorithms for connected dominating sets.
\newblock {\em Algorithmica}, 20(4):374--387, 1998.

\bibitem{GuhaK99}
S.~Guha and S.~Khuller.
\newblock Improved methods for approximating node weighted steiner trees and
  connected dominating sets.
\newblock {\em Information and Computation}, 150(1):57--74, 1999.

\bibitem{GuoGrammJCSS2006}
J.~Guo, J.~Gramm, F.~H{\"u}ffner, R.~Niedermeier, and S.~Wernicke.
\newblock Compression-based fixed-parameter algorithms for feedback vertex set
  and edge bipartization.
\newblock {\em Journal of Computer and System Sciences}, 72(8):1386--1396,
  2006.

\bibitem{GuoNW07}
J.~Guo, R.~Niedermeier, and S.~Wernicke.
\newblock Parameterized complexity of vertex cover variants.
\newblock {\em Theory of Computing Systems}, 41(3):501--520, 2007.

\bibitem{MR1192785}
F.~K. Hwang, D.~S. Richards, and P.~Winter.
\newblock {\em The {S}teiner tree problem}, volume~53 of {\em Annals of
  Discrete Mathematics}.
\newblock North-Holland Publishing Co., Amsterdam, 1992.

\bibitem{LokshtanovMS09}
D.~Lokshtanov, M.~Mnich, and S.~Saurabh.
\newblock Linear kernel for planar connected dominating set.
\newblock In {\em TAMC}, volume 5532 of {\em Lecture Notes in Computer
  Science}, pages 281--290, 2009.

\bibitem{LovaszGMT}
L.~Lov\'{a}sz.
\newblock Graph minor theory.
\newblock {\em Bulletin of the American Mathematical Society}, 43(1):75--86,
  2005.

\bibitem{MolleRR08}
D.~M{\"o}lle, S.~Richter, and P.~Rossmanith.
\newblock Enumerate and expand: Improved algorithms for connected vertex cover
  and tree cover.
\newblock {\em Theory of Computing Systems}, 43(2):234--253, 2008.

\bibitem{Nederlof2009}
J.~Nederlof.
\newblock Fast polynomial-space algorithms using m{\"o}bius inversion:
  Improving on steiner tree and related problems.
\newblock In {\em Proceedings of the 36th International Colloquium on Automata,
  Languages and Programming (ICALP 2009)}, pages 713--725, 2009.

\bibitem{Niedermeierbook06}
R.~Niedermeier.
\newblock {\em Invitation to Fixed-Parameter Algorithms}, volume~31 of {\em
  Oxford Lecture Series in Mathematics and its Applications}.
\newblock Oxford University Press, Oxford, 2006.

\bibitem{RamanSS2002}
V.~Raman, S.~Saurabh, and C.~Subramanian.
\newblock Faster fixed parameter tractable algorithms for finding feedback
  vertex sets.
\newblock {\em ACM Transactions on Algorithms}, 2(3):403--415, 2006.

\bibitem{RobinsZelikovsky2005}
G.~Robins and A.~Zelikovsky.
\newblock Tighter bounds for graph steiner tree approximation.
\newblock {\em SIAM Journal Discrete Mathematics}, 19(1):122--134, 2005.

\bibitem{SittersG2009}
R.~Sitters and A.~Grigoriev.
\newblock Connected feedback vertex set in planar graphs.
\newblock In {\em Proceedings of the 35th International Workshop on
  Graph-Theoretic Concepts in Computer Science (WG2009)}, 2009.

\bibitem{Thomasse2009}
S.~Thomass\'{e}.
\newblock A quadratic kernel for feedback vertex set.
\newblock In {\em Proceedings of the 19th Annual ACM-SIAM Symposium on Discrete
  Algorithms (SODA 2009)}, pages 115--119. Society for Industrial and Applied
  Mathematics, 2009.

\end{thebibliography}
